\documentclass[aps,prl,twocolumn,superscriptaddress]{revtex4}
\usepackage{amsmath}
\usepackage{amssymb}
\usepackage{hyperref}
\usepackage{url}
\usepackage{graphicx}
\usepackage{dcolumn} 
\usepackage{bm}      
\usepackage{multirow}
\usepackage{epstopdf}
\usepackage{color}
\usepackage{gensymb}
\usepackage{textcomp}
\usepackage{chemformula} 

\begin{document}

\title{Magnetic order and spin liquid behavior in [Mo$_3$]$^{11+}$ molecular magnets}

\author{Q.~Chen}
\affiliation{Department of Physics and Astronomy, University of Tennessee,
Knoxville, Tennessee 37996-1200, USA}

\author{R.~Sinclair}
\affiliation{Department of Physics and Astronomy, University of Tennessee,
Knoxville, Tennessee 37996-1200, USA}

\author{A.~Akbari-Sharbaf}
\affiliation{Institut Quantique and Département de Physique, Université de Sherbrooke, 2500 boul. de l'Université, Sherbrooke (Québec) J1K 2R1 Canada}

\author{Q.~Huang}
\affiliation{Department of Physics and Astronomy, University of Tennessee,
Knoxville, Tennessee 37996-1200, USA}

\author{Z.~Dun}
\affiliation{School of Physics, Georgia Institute of Technology, Atlanta, Georgia 30332, USA}

\author{E.~S.~Choi}
\affiliation{High Magnetic Field Laboratory, Florida State University, Tallahassee, Florida 32306-4005, USA} 

\author{M.~Mourigal}
\affiliation{School of Physics, Georgia Institute of Technology, Atlanta, Georgia 30332, USA}

\author{A.~Verrier}
\author{R.~Rouane}
\author{X.~Bazier-Matte}

\author{J.~A.~Quilliam}
\email[Email to: ]{jeffrey.quilliam@usherbrooke.ca}
\affiliation{Institut Quantique and Département de Physique, Université de Sherbrooke, 2500 boul. de l'Université, Sherbrooke (Québec) J1K 2R1 Canada}

\author{A.~A.~Aczel}
\email[Email to: ]{aczelaa@ornl.gov}
\affiliation{Neutron Scattering Division, Oak Ridge National Laboratory, Oak Ridge, Tennessee 37831, USA}

\author{H.~D.~Zhou}
\email[Email to: ]{hzhou10@utk.edu}
\affiliation{Department of Physics and Astronomy, University of Tennessee,
Knoxville, Tennessee 37996-1200, USA}

\date{\today}

\begin{abstract}
Molecular magnets based on [Mo$_3$]$^{11+}$ units with one unpaired electron per trimer have attracted recent interest due to the identification of quantum spin liquid candidacy in some family members. Here, we present comprehensive measurements on polycrystalline samples of \ch{ZnScMo3O8}, \ch{MgScMo3O8}, and \ch{Na3Sc2Mo5O16} with the same Mo$_3$O$_{13}$ magnetic building blocks. The crystal structures are characterized with x-ray or neutron powder diffraction and the magnetic ground states are determined by performing ac \& dc susceptibility, specific heat, neutron powder diffraction, and $\mu$SR measurements. Our work indicates that \ch{ZnScMo3O8} and \ch{MgScMo3O8} have ferromagnetic Curie-Weiss temperatures of 18.5 K and 11.9 K, ordered ground states with net moments (low-moment ferromagnetism or canted antiferromagnetism), and zero field ordering temperatures of $T_c =$~6~K and $<$~2~K respectively. On the other hand, \ch{Na3Sc2Mo5O16} hosts a dynamical magnetic ground state with no evidence for magnetic ordering or spin freezing down to 20 mK despite an antiferromagnetic Curie-Weiss temperature of -36.2 K, and therefore is a candidate for quantum spin liquid behavior. By comparing the present results to past work on the same family of materials, we construct a phase diagram which illustrates that the magnetic ground states of these Mo-based molecular magnets are very sensitive to small changes in the nearest neighbor Mo-Mo distance. 
\end{abstract}

\maketitle

\section{I. INTRODUCTION}
A quantum spin liquid (QSL) is a dynamical state of matter characterized by long-range entanglement between spins and fractionalized quasiparticle excitations \cite{73_anderson, 10_balents, 17_savary, 19_knolle, 19_wen, 20_broholm, 21_chamorro} such as spinons \cite{13_mourigal, 16_wu} or Majorana fermions \cite{16_banerjee, 17_banerjee, 19_takagi}. There is immense interest in identifying and characterizing QSL states in two and three-dimensional magnets because only their one-dimensional analogs have been unambiguously identified to-date \cite{20_broholm} and some QSLs may play an important role in future quantum computers \cite{05_kitaev, 10_bonderson, 11_alicea}. The typical ingredients in QSL candidates are small spins and competing interactions (i.e. geometric frustration \cite{94_ramirez}), hence many researchers have focused on Cu$^{2+}$ with spin-1/2 on the Kagome \cite{07_mendels, 09_okamoto, 11_kermarrec, 12_han, 20_smaha} and triangular lattices \cite{08_itou, 11_zhou, 12_quilliam}. One can also generate an effective spin-1/2 degree of freedom via strong spin-orbit coupling in Co$^{2+}$ \cite{78_tellenbach, 83_nagler, 17_ross} or Ir$^{4+}$ \cite{08_kim} octahedrally-coordinated magnets or with a crystal field ground state doublet well separated from excited states in rare-earth ions like Yb$^{3+}$ or Ce$^{3+}$ \cite{15_gaudet, 16_hallas, 19_gao, 19_gaudet, 20_sibille}. These two paradigms have generated a slew of interesting QSL candidates on the pyrochlore lattice (e.g., Yb$_2$Ti$_2$O$_7$ \cite{11_ross}, Ce$_2$Zr$_2$O$_7$ \cite{19_gao, 19_gaudet}), the triangular lattice (e.g., YbMgGaO$_4$ \cite{16_shen} and NaYbO$_2$ \cite{19_bordelon, 20_bordelon}), and the honeycomb lattice (e.g., $\alpha$-RuCl$_3$ \cite{16_banerjee, 17_banerjee}). Notably, the frustration in the honeycomb geometry arises from competing bond-directional Ising interactions and gives rise to the Kitaev model. This Hamiltonian is exactly solvable and yields a Kitaev spin liquid ground state with fractionalized Majorana fermion excitations \cite{06_kitaev}.  

\begin{figure}[tbp]
\linespread{1}
\par
\begin{center}
\includegraphics[width=\columnwidth]{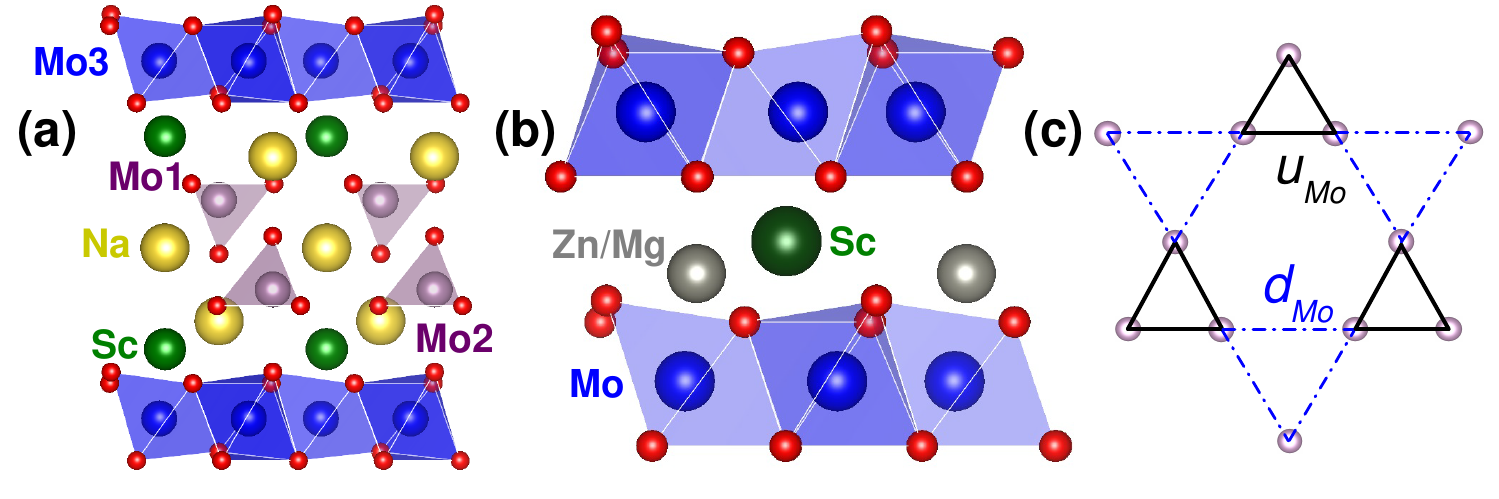}
\end{center}
\par
\caption{(color online) (a) The crystal structure of Na$_3$Sc$_2$Mo$_5$O$_{16}$. (b) The crystal structure of (Zn/Mg)ScMo$_3$O$_8$. (c) A schematic of the breathing kagome lattice formed by the Mo atoms in the $ab$-plane, with the small up triangles and large down triangles indicated by the solid black lines and the blue dash-dotted lines respectively. The Mo-Mo bond lengths are labelled $u_{Mo}$ and $d_{Mo}$ for the up and down triangles respectively. }
\label{Fig1}
\end{figure}

More recently, it has been appreciated that molecular magnets can also generate interacting spin-1/2 degrees of freedom on frustrated lattice geometries \cite{00_pocha, 12_sheckelton, 17_ziat}, leading to novel collective magnetic and electronic phenomena \cite{04_abdelmeguid, 14_haraguchi, 12_sheckelton, 17_sheckelton, 17_haraguchi, 18_akbari}. Molecular magnets are constructed from molecular building blocks with extremely short metal-metal distances. This ensures that some valence electrons participate in metal-metal bonding, while the remaining unpaired electrons provide the spin degree of freedom. A prime example of this behavior was identified in LiZn$_2$Mo$_3$O$_8$, which consists of [Mo$_3$]$^{11+}$ trimers \cite{85_torardi}. There are seven valence electrons per cluster. Six of these electrons generate metal-metal bonds to hold the cluster together, while the remaining unpaired electron provides the spin-1/2 per trimer. The [Mo$_3$]$^{11+}$ clusters are arranged in planes that are perpendicular to the c-axis and separated by non-magnetic spacer layers, as shown in Fig.~\ref{Fig1}(a) and (b). 

Initial characterization of the magnetic ground state for LiZn$_2$Mo$_3$O$_8$ showed that 2/3 of the spins condense into dynamic spin singlets, while the remaining 1/3 remain paramagnetic down to low temperatures \cite{12_sheckelton}. Subsequent theoretical work suggested that the spin-1/2 magnetic sublattice for this material may be best described by a 1/6th-filled breathing kagome network due to large intracluster and intercluster Coulomb repulsions \cite{16_chen, 21_nikolaev}. As shown in Fig.~\ref{Fig1}(c), this lattice geometry consists of small ``up" triangles and large ``down" triangles with Mo-Mo bond lengths of $u_\mathrm{Mo}$ and $d_\mathrm{Mo}$ respectively and a spin-1/2 moment occupying one vertex of each up triangle \cite{13_clark, 17_schaffer, 17_orain}. Within the context of a Hubbard model description, there is a specific parameter space that gives rise to a dynamic plaquette charge order \cite{16_chen, 21_nikolaev} and there are reasons to expect that this can lead to a gapless quantum spin liquid state with fractionalized spinon excitations \cite{16_chen}. Notably, the proposed model is consistent with the experimental observation that only 1/3 of the magnetic moments are active at low temperatures. Some circumstantial evidence for the quantum spin liquid ground state has also been found, including a gapless magnetic excitation spectrum in inelastic neutron scattering \cite{14_mourigal}, no evidence for static magnetism in muon spin relaxation \cite{14_sheckelton}, and an increasing $1/T_1 T$ with decreasing temperature in nuclear magnetic resonance \cite{14_sheckelton}. 

Since the discovery of the exotic magnetic ground state of LiZn$_2$Mo$_3$O$_8$, other [Mo$_3$]$^{11+}$ molecular magnets with similar separation between the magnetic planes have been identified and characterized. Li$_2$InMo$_3$O$_8$ hosts 120$^\circ$ long-range antiferromagnetic order \cite{13_gall, 15_haraguchi, 18_akbari, 19_iida} whereas the ground state of Li$_2$ScMo$_3$O$_8$ consists of coexisting regions of static and dynamic magnetism \cite{15_haraguchi, 18_akbari, 19_iida}. For intermediate stoichiometries in between those of the two parent compounds, for example in Li$_2$In$_{0.4}$Sc$_{0.6}$Mo$_3$O$_8$ \cite{18_akbari}, a dynamical magnetic ground state has been discovered. This system is therefore a quantum spin liquid candidate and has similarities to Li$_2$ZnMo$_3$O$_8$, including a crossover temperature below which 2/3 of the magnetic moments are lost and the Curie constant has a reduced value \cite{18_akbari}. There are also variants with more complicated spacer layers as shown in Fig.~\ref{Fig1}(a), such as Na$_3$Sc$_2$Mo$_5$O$_{16}$ and Na$_3$In$_2$Mo$_5$O$_{16}$ \cite{17_haraguchi_2}, that cause the magnetic [Mo$_3$]$^{11+}$ planes to be spaced much further apart. Magnetic susceptibility and heat capacity measurements find no evidence for long-range magnetic order in these two materials, which may be due to significantly weaker interplane exchange interactions or the realization of a quantum spin liquid ground state. 

Motivated by the high tunability of the magnetic ground states and the potential for uncovering exotic phases in these [Mo$_3$]$^{11+}$ molecular magnets, in this work we investigate the magnetic properties of \ch{ZnScMo3O8}, \ch{MgScMo3O8}, and Na$_3$Sc$_2$Mo$_5$O$_{16}$ through a combination of magnetic susceptibility, specific heat, neutron powder diffraction, and muon spin relaxation measurements. Notably, we extend the magnetic characterization of Na$_3$Sc$_2$Mo$_5$O$_{16}$ to dilution fridge temperatures. We find that \ch{ZnScMo3O8} and \ch{MgScMo3O8} have ordered ground states with net moments and Na$_3$Sc$_2$Mo$_5$O$_{16}$ shows no sign of magnetic order down to 20~mK, which provides additional support for its quantum spin liquid candidacy. We also compare our present results to previous work on the [Mo$_3$]$^{11+}$ molecular magnets in a phase diagram to illustrate the important role that the breathing parameter $\lambda_B = d_{Mo}/u_{Mo}$ and the nearest neighbor Mo-Mo distance $u_{Mo}$ play in the evolution of the magnetic ground states for this family of materials. 

\section{II. EXPERIMENTAL METHODS}

Polycrystalline samples of ZnScMo$_3$O$_8$, MgScMo$_3$O$_8$, and Na$_3$Sc$_2$Mo$_5$O$_{16}$ were synthesized by solid state reactions. For ZnScMo$_3$O$_8$ and MgScMo$_3$O$_8$, initially a stoichiometric mixture of ZnO/MgO, Mo, MoO$_3$, and Sc$_2$O$_3$ was ground together, pressed, sealed in an evacuated quartz tube, and then annealed at 1000 $\degree$C and 1150 $\degree$C for 48 hours and 24 hours respectively. Although the ZnScMo$_3$O$_8$ sample contained a large amount of MoO$_2$ impurity ($>$~10~wt\%), we found that including an additional 25~wt\% ZnO in the starting mixture was sufficient to completely suppress it. A similar issue was identified for the MgScMo$_3$O$_8$ sample, although in this case it took an additional 50~wt\% MgO to completely suppress the MoO$_2$ impurity. For Na$_3$Sc$_2$Mo$_5$O$_{16}$, Na$_2$MoO$_4$ was first synthesized by annealing a pellet consisting of a stoichiometric mixture of Na$_2$CO$_3$ and MoO$_3$ at 600 $\degree$C for 20 hours in air. We then ground together a stoichiometric mixture of Na$_2$MoO$_4$, Mo, MoO$_3$, and Sc$_2$O$_3$, pressed it into a pellet, sealed it in an evacuated quartz tube, and then annealed the pellet at 700 $\degree$C for 48 hours.

Room temperature x-ray diffraction (XRD) patterns were recorded with a HUBER Imaging Plate Guinier Camera 670 using Cu $K_{\alpha1}$ radiation (1.54~\AA). Complementary high-resolution neutron powder diffraction (NPD) measurements were performed using the HB-2A powder diffractometer at the High Flux Isotope Reactor (HFIR) of Oak Ridge National Laboratory with samples consisting of several grams of ZnScMo$_3$O$_8$ or Na$_3$Sc$_2$Mo$_5$O$_{16}$. We used a Ge monochromator to select neutron wavelengths of 1.54 and 2.41 {\AA} and a collimation of open-$21'$-$12'$. The ZnScMo$_3$O$_8$ (Na$_3$Sc$_2$Mo$_5$O$_{16}$) sample was loaded in a cylindrical V (Cu) can, which was then mounted in a cryostat (dilution fridge insert of a cryostat). These two different setups enabled the ZnScMo$_3$O$_8$ sample to achieve a base temperature of 1.5 K and the Na$_3$Sc$_2$Mo$_5$O$_{16}$ sample to achieve a base temperature of 50 mK. The XRD and HB-2A NPD data refinements were performed using the software package FULLPROF \cite{93_rodriguez}. Elastic neutron scattering measurements were also conducted on a 6~g polycrystalline sample of ZnScMo$_3$O$_8$ using the 14.6~meV fixed-incident-energy triple-axis spectrometer HB-1A of the HFIR at ORNL. Since the main goal of this experiment was to look for weak antiferromagnetic Bragg peaks, the sample was loaded in an Al can to reduce incoherent elastic scattering. The background was also minimized by using a double-bounce pyrolytic graphite (PG) monochromator system, mounting two-highly oriented PG filters in the incident beam to remove higher-order wavelength contamination, and placing a PG analyzer crystal before the single He-3 detector for energy discrimination. A collimation of 40$'$-40$'$-40$'$-80$'$-open resulted in an energy resolution at the elastic line just over 1 meV (FWHM). 

The dc magnetic susceptibility measurements were performed using a Quantum Design Superconducting Quantum Interference Device (SQUID) magnetometer. The specific heat measurements were performed using a Quantum Design Physical Property Measurement System (PPMS). For ZnScMo$_3$O$_8$, the high-field dc magnetization was measured using a vibrating sample magnetometer (VSM) and the ac susceptibility was measured with the conventional mutual inductance technique using a homemade setup \cite{14_dun} at the National High Magnetic Field Laboratory. For MgScMo$_3$O$_8$, the dc magnetization was measured up to 14~T using a Quantum Design PPMS. 

Muon spin relaxation ($\mu$SR) experiments were performed at TRIUMF in zero field (ZF) and longitudinal fields $B_L$. All three samples were studied on the M20 beam line using the LAMPF spectrometer with a low-background ``veto'' set-up, where they were suspended in mylar packets in a helium-flow cryostat within the path of the muon beam. This configuration allowed for a base temperature of roughly 2 K and longitudinal fields as high as 0.2~T. For Na$_3$Sc$_2$Mo$_5$O$_{16}$, additional measurements were performed on the M15 beam line with the sample fixed to the silver cold finger of a dilution refrigerator to achieve temperatures between $\sim 20$ mK and 3~K. A weakly-relaxing background term arising from this setup was subtracted off by comparing this data with M20 data collected in an overlapping temperature range. Longitudinal fields as high as 1~T were applied at low temperatures.

\section{III. RESULTS AND DISCUSSION}

\subsection{Na$_3$Sc$_2$Mo$_5$O$_{16}$}

\begin{figure}[tbp]
\linespread{1}
\par
\begin{center}
\includegraphics[width=\columnwidth]{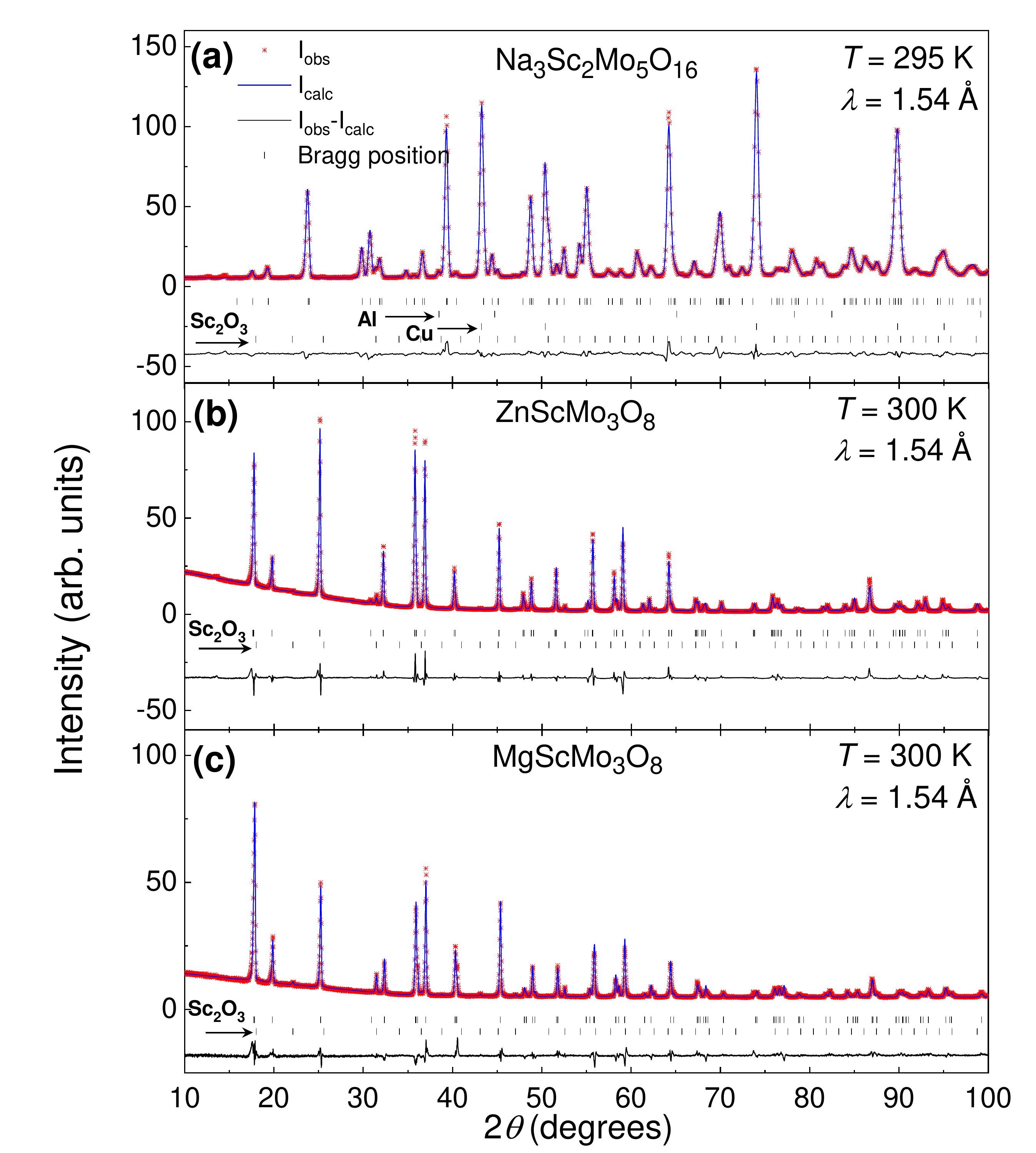}
\end{center}
\par
\caption{(color online) The powder diffraction patterns (red crosses) with $\lambda =$~1.54~\AA~for polycrystalline (a) Na$_3$Sc$_2$Mo$_5$O$_{16}$ (NPD, 295 K), (b) ZnScMo$_3$O$_8$ (XRD, room temperature), and (c) MgScMo$_3$O$_8$ (XRD, room temperature). The solid blue curves are the best fits from the Rietveld refinements using FULLPROF. The vertical tick marks indicate the position of Bragg reflections for the [Mo$_3$]$^{11+}$ molecular magnet, the Sc$_2$O$_3$ impurity phase, or Cu and Al background contributions. The solid black curves show the difference between the observed and calculated intensities.}
\label{Fig2}
\end{figure}

Although the room-temperature XRD data was useful for confirming that the Na$_3$Sc$_2$Mo$_5$O$_{16}$ sample crystallized in the expected $P3m1$ space group, a complete refinement was not possible due to the large number of oxygen parameters in the structure. For this reason, we also collected the HB-2A neutron powder diffraction data at $T =$~295~K with $\lambda =$~1.54~\AA~that is presented in Fig.~\ref{Fig2}(a). The Rietveld refinement yields the lattice parameters $a=b=$~5.7942(2)~\AA~and $c=$~11.1341(4)~\AA, which are slightly larger than the previously reported values of $a=b=$~5.78373(4)~\AA~ and $c=$~11.1078(1)~\AA~\cite{17_haraguchi_2}. Despite this small difference in the lattice constants, the breathing parameter $\lambda_B =$~1.24 is nearly identical to the value obtained in the previous work, while the nearest neighbor Mo-Mo distance of 2.59~\AA~is slightly larger. Detailed crystallographic parameters from the NPD refinement are listed in Table~\ref{Table I}. There are some small impurity peaks in this data that can be attributed to 1.4~wt\% non-magnetic Sc$_2$O$_3$. There are also two additional small impurity peaks between 2$\theta =$~12 and 15 degrees that could not be indexed. 

\begin{table*}[]
\centering
\caption{Structural parameters for (a) Na$_3$Sc$_2$Mo$_5$O$_{16}$, (b) ZnScMo$_3$O$_8$, and (c) MgScMo$_3$O$_8$, obtained from refinements of powder diffraction data.}
\label{Table I}
\setlength{\tabcolsep}{0.5 cm}
\begin{tabular}{ccccccc}
\hline \hline
Refinement & Atom & Wyckoff & \textit{x} & \textit{y} & \textit{z} & Occupancy \\ 
\hline
\multirow{7}{*}{\begin{tabular}[c]{@{}c@{}}\\ \\ \\ \\ \\ (a)\\Na$_3$Sc$_2$Mo$_5$O$_{16}$\\ NPD \\ $T =$~295~K \\ $\textit{R}_\text{wp}$ = 5.11\%\\ ($P3m1$) \end{tabular}} 
 & Sc1 & 1c & 2/3 & 1/3 & 0.194(1) & 0.16667 \\
 & Sc2 & 1c & 2/3 & 1/3 & 0.7893(9) & 0.16667 \\
 & Mo1 & 1b & 0 & 0 & 0.360(1) & 0.16667 \\
 & Mo2 & 1a & 1/3 & 2/3 & 0.611(1) & 0.16667 \\
 & Mo3 & 3d & 0.1489(5) & 0.8511(5) & 0 & 0.50000 \\
 & Na1 & 1c & 2/3 & 1/3 & 0.476(2) & 0.16667 \\
 & Na2 & 1a & 1/3 & 2/3 & 0.273(2) & 0.16667 \\
 & Na3 & 1b & 0 & 0 & 0.733(2) & 0.16667 \\
 & O1 & 3d & 0.842(1) & 0.158(1) & 0.882(1) & 0.50000 \\
 & O2 & 3d & 0.503(1) & 0.497(1) & 0.656(1) & 0.50000 \\
 & O3 & 1b & 0 & 0 & 0.510(2) & 0.16667 \\
 & O4 & 1a & 1/3 & 2/3 & 0.906(1) & 0.16667 \\
 & O5 & 3d & 0.495(1) & 0.505(1) & 0.098(1) & 0.50000 \\
 & O6 & 3d & 0.827(1) & 0.173(1) & 0.315(1) & 0.50000 \\
 & O7 & 1a & 1/3 & 2/3 & 0.460(2) & 0.16667 \\
 & O8 & 1b & 0 & 0 & 0.130(2) & 0.16667 \\
\multicolumn{1}{l}{} & \multicolumn{1}{l}{} & \multicolumn{1}{l}{} & \multicolumn{1}{l}{} & \multicolumn{1}{l}{} & \multicolumn{1}{l}{} & \multicolumn{1}{l}{} \\
 & \multicolumn{6}{c}{$a=b$ = 5.7942(2) \AA, $c$ = 11.1341(4) \AA} \\
\multicolumn{1}{l}{} & \multicolumn{6}{l}{} \\
\multicolumn{1}{l}{} & \multicolumn{6}{c}{Overall B-factor = 0.46(2) \AA$^{2}$} \\ 
\hline
\multirow{7}{*}{\begin{tabular}[c]{@{}c@{}}(b)\\ ZnScMo$_3$O$_8$\\ XRD \\ $T =$~300~K \\ $\textit{R}_\text{wp}$ = 7.71\%\\ ($P6_{3}mc$) \end{tabular}} 
 & Zn & 2b & 1/3 & 2/3 & 0.9532(2) & 0.16667 \\
 & Sc & 2b & 1/3 & 2/3 & 0.5098(3) & 0.16667 \\
 & Mo & 6c & 0.14489(6) & -0.14489(6) & 0.25 & 0.50000 \\
 & O1 & 2a & 0 & 0 & 0.3950(9) & 0.16667 \\
 & O2 & 2b & 1/3 & 2/3 & 0.1389(9) & 0.16667 \\
 & O3 & 6c & 0.4898(1) & 0.5102(1) & 0.3724(5) & 0.50000 \\
 & O4 & 6c & 0.16451(7) & -0.16451(7) & 0.6310(5) & 0.50000 \\
\multicolumn{1}{l}{} & \multicolumn{1}{l}{} & \multicolumn{1}{l}{} & \multicolumn{1}{l}{} & \multicolumn{1}{l}{} & \multicolumn{1}{l}{} & \multicolumn{1}{l}{} \\
 & \multicolumn{6}{c}{$a=b$ = 5.7983(1) \AA, $c$ = 9.9881(1) \AA} \\
\multicolumn{1}{l}{} & \multicolumn{6}{l}{} \\
\multicolumn{1}{l}{} & \multicolumn{6}{c}{Overall B-factor = 0.330(1) \AA$^{2}$} \\ 
\hline 
\multirow{7}{*}{\begin{tabular}[c]{@{}c@{}}(c)\\ MgScMo$_3$O$_8$\\ XRD \\ $T =$~300~K \\ $\textit{R}_\text{wp}$ = 5.87\%\\ ($P6_{3}mc$) \end{tabular}} 
 & Mg & 2b & 1/3 & 2/3 & 0.9218(6) & 0.16667 \\
 & Sc & 2b & 1/3 & 2/3 & 0.5042(4) & 0.16667 \\
 & Mo & 6c & 0.14387(7) & -0.14387(7) & 0.25 & 0.50000 \\
 & O1 & 2a & 0 & 0 & 0.429(1) & 0.16667 \\
 & O2 & 2b & 1/3 & 2/3 & 0.146(1) & 0.16667 \\
 & O3 & 6c & 0.4918(4) & 0.5082(4) & 0.3644(6) & 0.50000 \\
 & O4 & 6c & 0.1660(6) & -0.1660(6) & 0.6362(5) & 0.50000 \\
 &  &  &  &  &  &  \\
 & \multicolumn{6}{c}{$a=b$ = 5.7804(1) \AA, $c$ = 9.9472(2) \AA} \\
 &  &  &  &  &  &  \\
 & \multicolumn{6}{c}{Overall B-factor = 0.43(2) \AA$^{2}$} \\
\hline \hline
\end{tabular}
\end{table*}

\begin{figure}[tbp]
\linespread{1}
\par
\begin{center}
\includegraphics[width=\columnwidth]{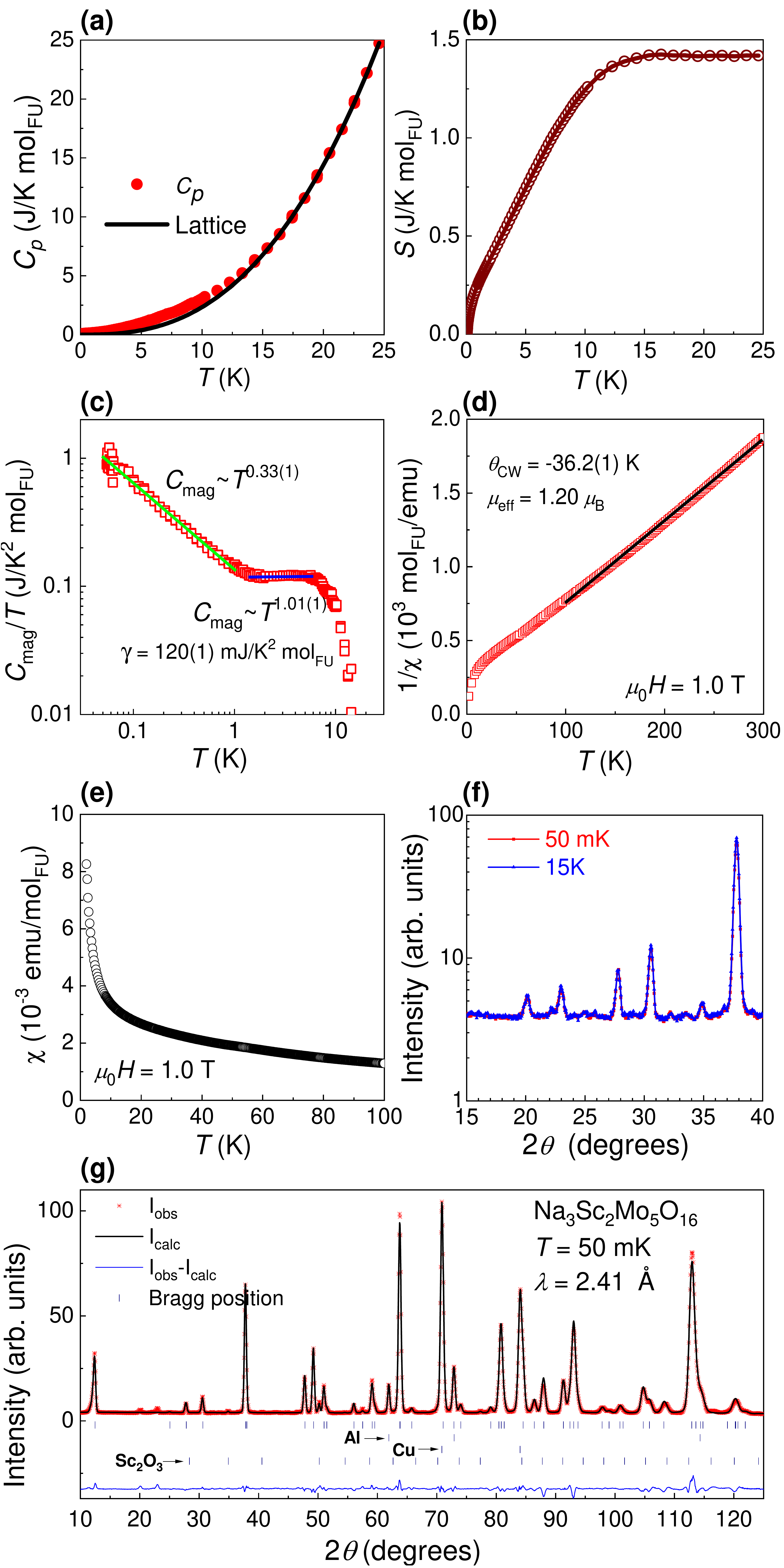}
\end{center}
\par
\caption{(color online) For Na$_3$Sc$_2$Mo$_5$O$_{16}$, 
(a) the $T$-dependence of the specific heat with the $T^3$ lattice contribution indicated by the solid curve,
(b) the magnetic entropy released when warming up to 25~K,
(c) the $T$-dependence of the magnetic contribution to the specific heat divided by temperature with power law fits (solid lines) superimposed on the data;
(d) the $T$-dependence of the inverse dc susceptibility and the associated Curie-Weiss fit (solid line);
(e) the $T$-dependence of the dc susceptibility over the low-$T$ region only;
(f) a comparison of the HB-2A neutron powder diffraction patterns for low angles only at $T =$~50~mK and 15~K;
(g) the full HB-2A neutron powder diffraction pattern at $T =$~50~mK, with the Rietveld refinement result (black curve) superimposed on the data. The vertical tick marks indicate the position of the Bragg reflections for the phase of interest, a small amount of Sc$_2$O$_3$ impurity, and background Al and Cu scattering, while the blue curve represents the fit residual.}
\label{Fig3}
\end{figure}

In this work we have extended the temperature range of the specific heat measurements for Na$_3$Sc$_2$Mo$_5$O$_{16}$ down to 60~mK. We isolated the magnetic contribution $C_{mag}$ to the specific heat by subtracting off the $T^3$ lattice contribution that was apparent in the higher-temperature regime, with the raw data and the $T^3$ contribution presented in Fig.~\ref{Fig3}(a). We also show the magnetic entropy that was obtained by integrating $C_{mag}/T$ in Fig.~\ref{Fig3}(b). As discussed in previous work \cite{17_haraguchi_2} and illustrated in Fig.~\ref{Fig3}(c), the magnetic contribution consists of two main features: a linear $T$-dependence ($C=\gamma T$) between 1 and 7 K and a sub-linear $T$-dependence at lower temperatures. The $\gamma$ value associated with the linear regime is 120(1)~mJ/K$^2$ mol$_{\rm FU}$, which is somewhat larger than previously reported but still a typical value for a gapless quantum spin liquid ground state \cite{08_yamashita, 11_yamashita, 11_cheng, 11_zhou}. The sub-linear regime at lower temperatures has already been attributed to a Schottky anomaly arising from a small number of free spins due to impurities and/or defects \cite{17_haraguchi_2}. 

The dc susceptibility ($\chi = M/H$) for Na$_3$Sc$_2$Mo$_5$O$_{16}$, plotted in Fig.~\ref{Fig3}(d) and (e) as 1/$\chi$ and $\chi$ respectively, also shows no sign of long range magnetic ordering down to 1.8 K. A Curie-Weiss fit of the high-temperature regime (between 100 and 300 K) for 1/$\chi$ yields a Curie-Weiss temperature $\theta_{\text{CW}}$ = -36.2(1)~K and an effective moment $\mu_{\text{eff}}$ = 1.20(3)~$\mu_{\text{B}}$/trimer. Although the magnitudes of these parameters are somewhat smaller than the values reported previously \cite{17_haraguchi_2}, they are still consistent with dominant antiferromagnetic interactions and an $S =$~1/2 moment for the [Mo$_3$]$^{11+}$ trimer units. The reduced Curie-Weiss temperature in our sample may be a consequence of the slightly larger lattice parameter in the breathing kagome plane.

To further corroborate the lack of magnetic order in Na$_3$Sc$_2$Mo$_5$O$_{16}$, we collected HB-2A neutron powder diffraction data down to 50~mK. Fig.~\ref{Fig3}(f) shows a comparison of the low-angle diffraction patterns for $T =$~50~mK and 15~K. No new magnetic Bragg peaks, which would be indicative of antiferromagnetic order, appear in the 50~mK data. There are also no visible intensity increases at the nuclear Bragg peak positions with decreasing temperature that would be expected for an ordered ground state with a net moment. The full HB-2A diffraction pattern at $T =$~50~mK, with the structural refinement superimposed on the data, is shown in Fig.~\ref{Fig3}(g). The space group remains $P3m1$ and the lattice parameters are $a =$~5.7857(1)~\AA~and $c = $~11.0970(2)~\AA. The two peaks between 2$\theta =$~20 and 25 degrees that are not indexed correspond to the same unidentified impurity phase described above.   

Finally, we performed a muon spin relaxation experiment to look for evidence of weak static magnetism that could not be detected by neutron diffraction. Asymmetry spectra in both zero field and applied longitudinal field, even at high temperatures, are well-described by double-exponential relaxation given by the following functional form:
\begin{equation}
P(t) = fe^{-\lambda_\mathrm{s}t} + (1-f) e^{-\lambda_\mathrm{f}t}, 
\label{DoubleExponential}\end{equation}
where $f =$~0.83 corresponds to the volume fraction of the slowly-relaxing component, $\lambda_s$ is the slow relaxation rate, and $\lambda_f$ is the fast relaxation rate. Here we have neglected the very weak Gaussian nuclear relaxation that should also be present in zero field. 

\begin{figure}[tbp]
\centering
\includegraphics[width=3.5in]{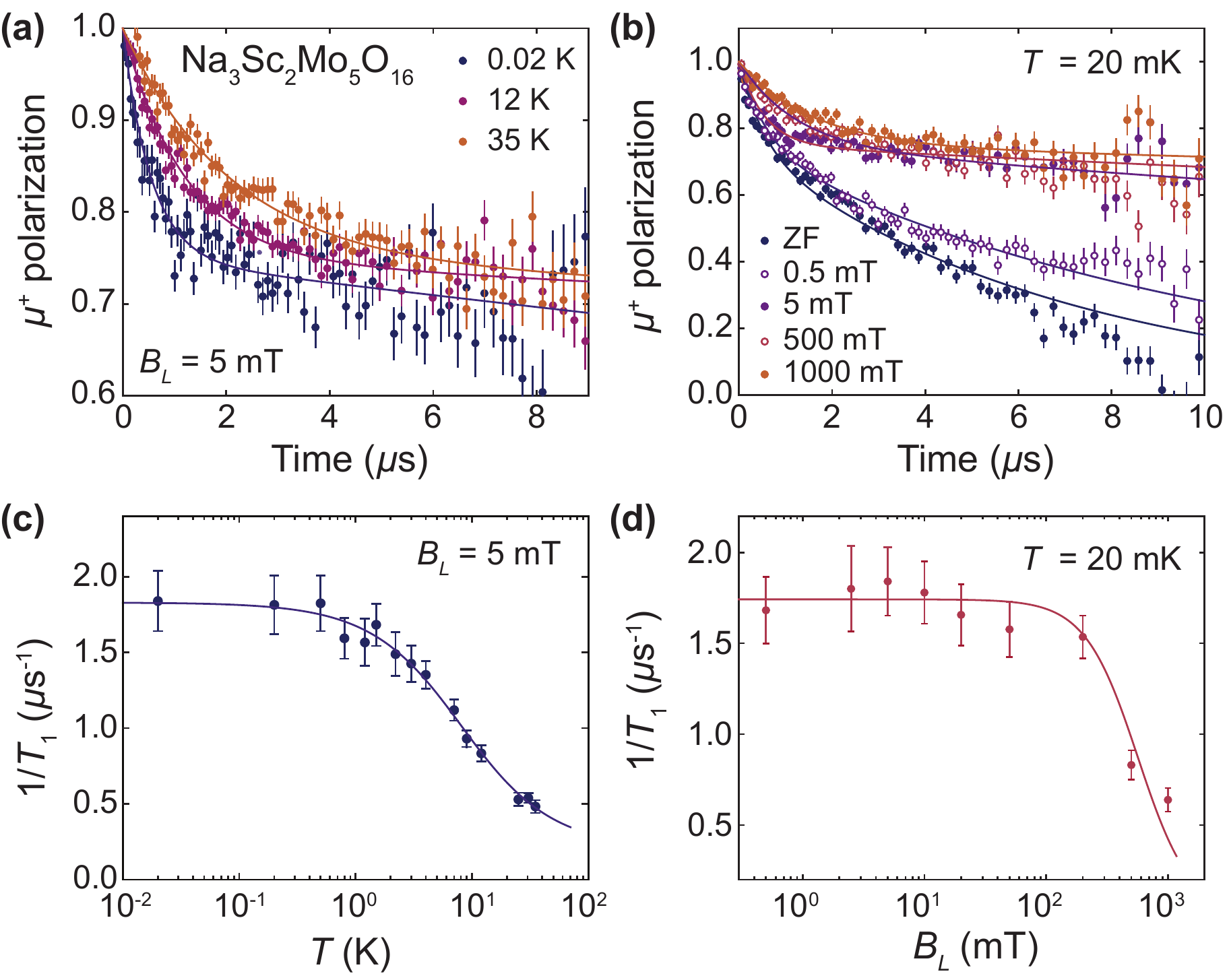}
\caption{(color online) (a) For Na$_3$Sc$_2$Mo$_5$O$_{16}$, muon spin polarization vs. time for selected temperatures with a longitudinal field of $B_L= 5$ mT. Fits to Eq.~(\ref{DoubleExponential}) are superimposed on the data. (b) Muon spin polarization vs time for selected longitudinal fields at $T =$~20~mK. Fits to Eq.~(\ref{DoubleExponential}) are superimposed on the data. (c) Relaxation rate of the fast-relaxing component $1/T_1$ as a function of temperature for $B_L = 5$ mT. The solid curve is a guide to the eye. (d) Longitudinal field-dependence of $1/T_1$ for the fast-relaxing component at $T=20$ mK with a Lorentzian fit (i.e. Redfield theory) superimposed on the data. 
\label{Fig4}}
\end{figure}

Figure~\ref{Fig4}(a) shows $B_L =$~5~mT longitudinal field spectra at selected temperatures. Application of such a weak longitudinal field is ideal for isolating the dynamic contribution to the relaxation $1/T_1$ since it completely decouples the muon spin polarization from small static fields that are nuclear in origin, while at the same time having little effect on the  relaxation that is caused by dynamical spin fluctuations, in particular the relaxation coming from the muons that are strongly coupled to the magnetic planes. There are no signs of oscillations or a sudden increase in relaxation rate that might indicate the development of static magnetism down to temperatures as low as 20 mK. Therefore, our $\mu$SR results are also supportive of a dynamical magnetic ground state for Na$_3$Sc$_2$Mo$_5$O$_{16}$. 

Since the two relaxation components are present at all temperatures, we hypothesize that they are likely a result of distinct $\mu^+$ stopping sites in the crystal structure. Positive muons are generally expected to stop roughly 1~\AA~away from the most electronegative ions in the material, in this case the oxygen ions \cite{83_holzschuh}. In this system, there are 8 distinct oxygen sites. Oxygens O1, O4, O5 and O8 form the magnetic MoO$_6$ octahedra and are at most $2.13$~\AA~away from the magnetic Mo3 sites. The remaining 4 oxygen sites form the non-magnetic MoO$_4$ tetrahedra in the spacer layer and are at least $3.87$~\AA~away from the Mo3 sites. Since the relaxation rate is proportional to $\Delta_D^2$, where $\Delta_D$ is the size of the fluctuating field~\cite{Slichter} and the dipolar interaction falls off as $1/r^3$, we can estimate that $\lambda_f/\lambda_s \simeq (3.87$ \AA $/ 2.13$ \AA$)^6 = 36$. At low temperatures, the measured ratio of the two relaxation rates, $\lambda_f/\lambda_s = 38$, agrees very well with this estimate and therefore lends credibility to our hypothesis.

As shown in Fig.~\ref{Fig4}(b), the muon spin polarization associated with the fast relaxation rate is not completely decoupled in longitudinal fields even as large as 1~T, a result that can only be obtained with relaxation that is dynamical in origin. Here we focus primarily on analyzing this component of the signal since changes in the slowly-relaxing term are difficult to resolve cleanly and are not as useful as a probe of the spin fluctuations of the magnetic Mo trimer planes. In Fig.~\ref{Fig4}(c), we show the fast relaxation rate, $1/T_1$, as a function of temperature at $B_L = 5$ mT. As with most other quantum spin liquid candidates \cite{07_mendels, 16_quilliam, 18_akbari}, this system shows a gradual increase in relaxation rate with decreasing temperature and a relaxation plateau below 1 K. Fig.~\ref{Fig4}(d) shows the same relaxation rate as a function of longitudinal field, which is fit with a single Lorentzian function of the form:
\begin{equation}
\frac{1}{T_1} = \frac{2 \gamma_\mu^2 \Delta_D^2 \omega_c}{\gamma_\mu^2 B_L^2 + \omega_c^2}, 
\label{Redfield}\end{equation}
where $\gamma_\mu = 2\pi \times$ 135.5 MHz/T is the muon gyromagnetic ratio, $\Delta_D$ is the magnitude of the fluctuating field, and $B_L$ is the applied longitudinal field. This expression is appropriate for fast fluctuating fields and it arises from Redfield theory for a single characteristic fluctuation rate $\omega_c$~\cite{Slichter}. The fit gives a fluctuating field of $\Delta_D = 24.1(1.4)$ mT and a characteristic frequency of $\omega_c = 480(60)$ MHz. 

\subsection{ZnScMo$_3$O$_8$}
The room temperature XRD pattern for ZnScMo$_3$O$_8$ shown in Fig.~\ref{Fig2}(b) confirms the expected hexagonal space group $P6_{3}mc$. A Rietveld refinement of this data yields lattice constants of $a =$~5.7983(1)~\AA~and $c =$~9.9881(1)~\AA, which are slightly smaller than the previously reported values of $a =$~5.8050(7)~\AA~and $c =$~9.996(3)~\AA~\cite{85_torardi}. A breathing parameter $\lambda_B =$~1.30 and a nearest neighbor Mo-Mo distance 2.52~\AA~are also obtained from the refinement. Detailed crystallographic parameters from the XRD refinement are listed in Table I. There are some small impurity peaks in this data that can be indexed to 3~wt\% non-magnetic Sc$_2$O$_3$.  
\begin{figure*}[tbp]
\linespread{1}
\par
\begin{center}
\includegraphics[width=7in]{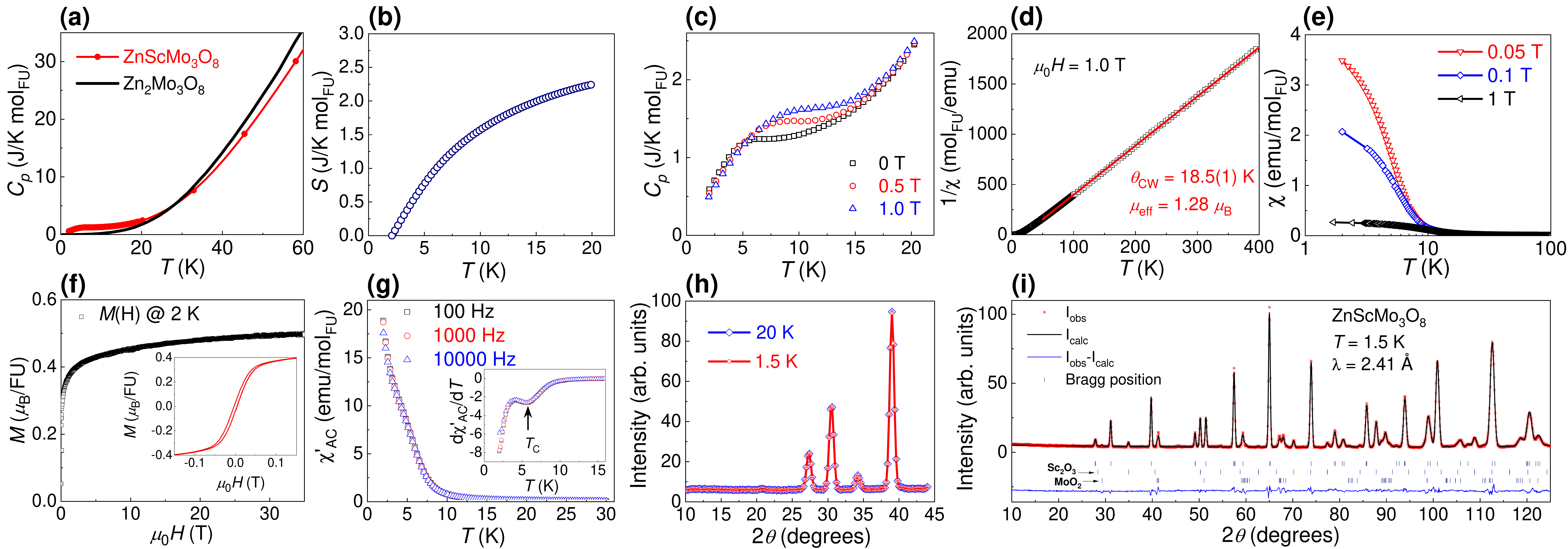}
\end{center}
\par
\caption{(color online) For ZnScMo$_3$O$_8$, 
(a) the $T$-dependence of the zero-field specific heat, along with data for the non-magnetic analog Zn$_2$Mo$_3$O$_8$;
(b) the magnetic entropy released when warming to 20 K;
(c) the $T$-dependence of the specific heat for selected magnetic fields;
(d) the $T$-dependence of the inverse dc susceptibility and the associated Curie-Weiss fit (solid line);
(e) the $T$-dependence of the dc susceptibility for selected fields over the low-$T$ region only;
(f) the magnetization $M$ vs applied field $H$ at $T =$~2~K, with the inset zoomed-in to the low-field regime where a small amount of hysteresis can be observed;
(g) the $T$-dependence of the real part of the ac susceptibility with an AC field amplitude of 5 Oe for various driving frequencies, with the first derivative plotted vs $T$ in the inset; 
(h) a comparison of the HB-1A neutron powder diffraction patterns for low angles only at $T =$~20~K and 1.5~K; 
(i) the full HB-2A neutron powder diffraction pattern at $T =$~1.5~K with the Rietveld refinement result (black curve) superimposed on the data. The vertical tick marks indicate the position of the Bragg reflections for ZnScMo$_3$O$_8$ and the MoO$_2$ and Sc$_2$O$_3$ impurities, while the blue curve represents the fit residual.}
\label{Fig5}
\end{figure*}

Although the crystal structure of ZnScMo$_3$O$_8$ has been determined previously, the magnetic properties were not reported. We measured the $T$-dependence of the specific heat down to 1.8~K, with the raw data shown in Fig.~\ref{Fig5}(a), and found a broad peak centered at 6~K. We isolated the magnetic signal by subtracting off the lattice contribution from isostructural, non-magnetic Zn$_2$Mo$_3$O$_8$. Although there is not a perfect lattice match in the higher-$T$ regime, it is clear that a magnetic contribution persists in the ZnScMo$_3$O$_8$ data well above the broad peak and becomes negligible between 20 and 25 K. The identification of a magnetic heat capacity signal well above a probable ordering transition is a hallmark of low-dimensional and/or frustrated magnetism. The magnetic entropy recovered by warming up to 20 K is presented in Fig.~\ref{Fig5}(b) and we find that the value is significantly lower than $Rln(2)$ expected for a magnetic ordering transition associated with a spin-1/2 system. This reduced value provides additional evidence for a significantly frustrated and/or low-dimensional lattice. We also collected specific heat data with selected applied fields, as illustrated in Fig.~\ref{Fig5}(c), and found that the broad peak shifts to higher temperatures as the field is increased. This behavior is consistent with a phase transition to an ordered state with a net moment.  

The dc susceptibility ($\chi = M/H$) for ZnScMo$_3$O$_8$, plotted in Fig.~\ref{Fig5}(d) and (e) as 1/$\chi$ and $\chi$ respectively, is also indicative of a magnetic ground state with a net moment. A Curie-Weiss fit of the high-temperature regime (between 50 and 400 K) for 1/$\chi$ yields a Curie-Weiss temperature $\theta_{\text{CW}}$ = 18.5(1)~K and an effective moment $\mu_{\text{eff}}$ = 1.28(3)~$\mu_{\text{B}}$/trimer. The positive Curie-Weiss temperature signifies dominant ferromagnetic interactions and the effective moment value implies that each [Mo$_3$]$^{11+}$ trimer unit carries $S =$~1/2. The $\chi$ data shows deviations from the Curie-Weiss law at low temperatures, with a sharp increase at low fields and the concave down behavior that is expected for magnetic materials with net moments. The magnetization vs field at 2~K, presented in Fig.~\ref{Fig5}(f), quantifies this net moment as it saturates at 0.5~$\mu_B$/trimer. There is very little change in the high-field magnetization up to 35~T, which indicates that there are no field-induced transitions in this regime. We also note that the saturation magnetization is well-below the expected value of 1~$\mu_B$/trimer for a spin-1/2 ferromagnet, which may be due to the close proximity of the measurement to the magnetic transition temperature, the importance of orbital angular momentum to the ground state, and/or a canted antiferromagnetic ground state. Finally, the ac susceptibility data shown in Fig.~\ref{Fig5}(g) reveals no frequency dependence in the low-$T$ data, as expected for a phase transition to long-range magnetic order. The first derivative of the ac susceptibility with respect to temperature is used to determine $T_c =$~6~K in the low-field limit. 

\begin{figure*}
\centering
\includegraphics[width=5.4in]{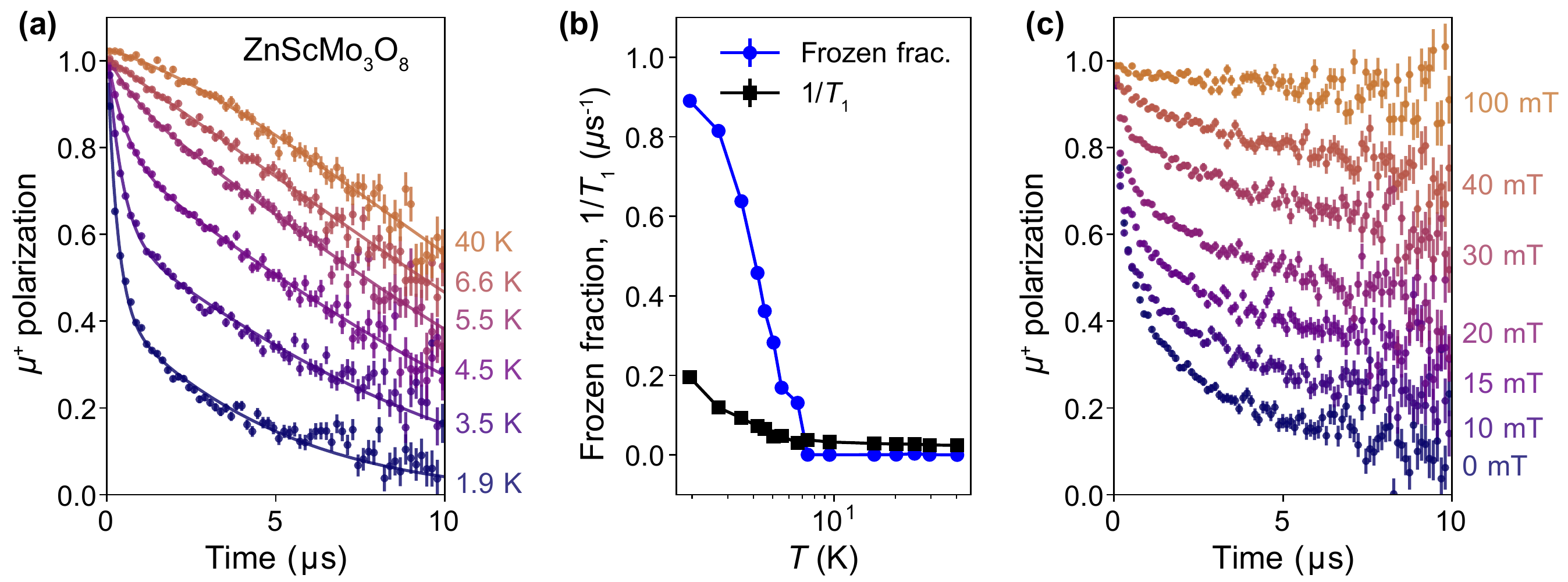}
\caption{(color online)
For ZnScMo$_3$O$_8$, (a) muon spin polarization vs. time in zero field for selected temperatures. Fits to a two-component function, Eq.~(\ref{ZnSc_fitting_function}), are superimposed on the data. Clear development of a fast-relaxing static component at low temperatures is observed.  (b) Temperature dependence of the ordered/frozen volume fraction $f$ and spin-lattice relaxation rate $1/T_1$. (c) Muon spin polarization vs. time for selected longitudinal fields at $T =$~1.95~K showing that the decoupling of the muon spin relaxation is consistent with largely static internal fields.
\label{Fig6}}
\end{figure*}

In an effort to differentiate between low-moment ferromagnetism and canted antiferromagnetism in ZnScMo$_3$O$_8$, we collected HB-1A neutron powder diffraction data down to 1.5~K to look for antiferromagnetic Bragg peaks. This experiment was performed on HB-1A rather than HB-2A due to the superior signal-to-noise ratio of the former instrument that arises from using a double-bounce PG monochromator system and a PG analyzer for energy discrimination. Fig.~\ref{Fig5}(h) shows a comparison of the low-angle diffraction patterns for $T =$~20~K and 1.5~K. There is no evidence for the new magnetic Bragg peaks at $T =$~1.5 K that would be expected for a canted antiferromagnetic ground state. There is also no detectable intensity enhancement of the low-angle nuclear Bragg peaks that would be indicative of an ordered ground state with a net moment. The lack of any magnetic signal in the HB-1A data prevents us from unambiguously differentiating between ferromagnetic and canted antiferromagnetic ground states, although this result confirms that the ordered moment size is quite small. Complementary HB-2A data was also collected at $T =$~1.5~K and the full diffraction pattern is plotted in Fig.~\ref{Fig5}(i), with the structural refinement superimposed on the data. The space group remains $P6_{3}mc$ and the lattice parameters are $a =$~5.7916(3)~\AA~and $c=$~9.9871(5)~\AA. No evidence for magnetic scattering intensity is visible in this data either. Note that the HB-2A data was collected on a ZnScMo$_3$O$_8$ sample prepared by a stoichiometric solid state reaction, so it contains a large amount of non-magnetic MoO$_2$ \cite{80_khilla, 00_eyert} impurity (13 wt\%) in addition to a small amount of Sc$_2$O$_3$ impurity (4 wt\%). 

$\mu$SR measurements were also carried out on ZnScMo$_3$O$_8$ to verify the magnetic ground state and assess the ordered volume fraction. The $T$-dependence of the ZF-$\mu$SR spectra is shown in Fig.~\ref{Fig6}(a). At high temperatures, we observe weak, $T$-independent Gaussian relaxation that typically arises from quasi-static nuclear moments. As the temperature decreases towards $T_c$, the spectra change drastically and a two-component signal develops. These data can be fit successfully with the following two-component function:
\begin{equation} P(t) = \{f(T)P_{\rm Static}(t) + [1-f(T)]P_{\rm Nuclear}(t)\} e^{-t/T_1} \label{ZnSc_fitting_function}
\end{equation}
where the static component is given by a fast exponential relaxation and 1/3-tail typical of magnetic order in powder samples
\begin{equation} P_{\rm Static}(t) = \frac{2}{3} e^{-\lambda_s} + \frac{1}{3}. \end{equation}
$P_{\rm Nuclear}(t)$ is the standard Gaussian Kubo-Toyabe function for nuclear magnetism. For simplicity, we have applied the same spin-lattice relaxation rate $1/T_1$ throughout the sample. As shown in Fig.~\ref{Fig6}(b), the frozen fraction $f(T)$ develops quite quickly at $T_c$ and is approaching a value close to 100\% at the lowest temperatures measured here. 
This indicates that we are observing the development of magnetic order in more or less the entire sample, although the simple exponential relaxation is likely indicative of a broadened local field distribution due to crystallographic defects or domain walls. Similar spectra were observed previously for the molecular magnets Ba$_3$InRu$_2$O$_9$ and Ba$_3$YRu$_2$O$_9$ \cite{17_ziat}, as well as the spin-1/2 kagome antiferromagnet vesignieite~\cite{11_colman}. The fast relaxation rate, $\lambda_s \simeq 3.4$ $\mu$s$^{-1}$ at 1.95 K, indicates an average internal field, $B_\mathrm{int} \simeq 4.0$ mT.

The effect of a longitudinal magnetic field on the relaxation is presented in Fig.~\ref{Fig6}(c) and shows that the muon spins decouple relatively quickly from the internal fields. Applied fields of 40 to 100 mT appear to be sufficient to achieve complete decoupling, thereby eliminating the fast-relaxing front-end of the spectra. Since these fields are on the order of $10B_\mathrm{int}$, this corroborates the hypothesis that this material is magnetically ordered and in the quasi-static limit at the lowest temperatures studied. 

\subsection{MgScMo$_3$O$_8$}
We synthesized MgScMo$_3$O$_8$ for the first time. The room temperature XRD pattern for the sample with an additional 50~wt\% MgO added in the synthesis process is consistent with the same hexagonal space group $P6_{3}mc$ identified for ZnScMo$_3$O$_8$. A Rietveld refinement of this data  yields lattice constants of $a =$~5.7804(1)~\AA~and $c =$~9.9472(2)~\AA, which are lower than the values for ZnScMo$_3$O$_8$ due to the smaller ionic radius of Mg$^{2+}$ relative to Zn$^{2+}$. A breathing parameter $\lambda_B =$~1.32 and a nearest neighbor Mo-Mo distance 2.49~\AA~are also obtained from the refinement. Detailed crystallographic parameters from the XRD refinement are listed in Table I. There are some small impurity peaks in this data that can be indexed to 8~wt\% non-magnetic Sc$_2$O$_3$. We note that the rest of the MgScMo$_3$O$_8$ data reported below was collected on a sample with only an additional 35~wt\% MgO added in the synthesis process so the non-magnetic MoO$_2$ impurity was not completely suppressed (about 4~wt\%), but this does not significantly affect our results. The Sc$_2$O$_3$ impurity content was essentially the same. 

\begin{figure}[tbp]
\linespread{1}
\par
\begin{center}
\includegraphics[width=\columnwidth]{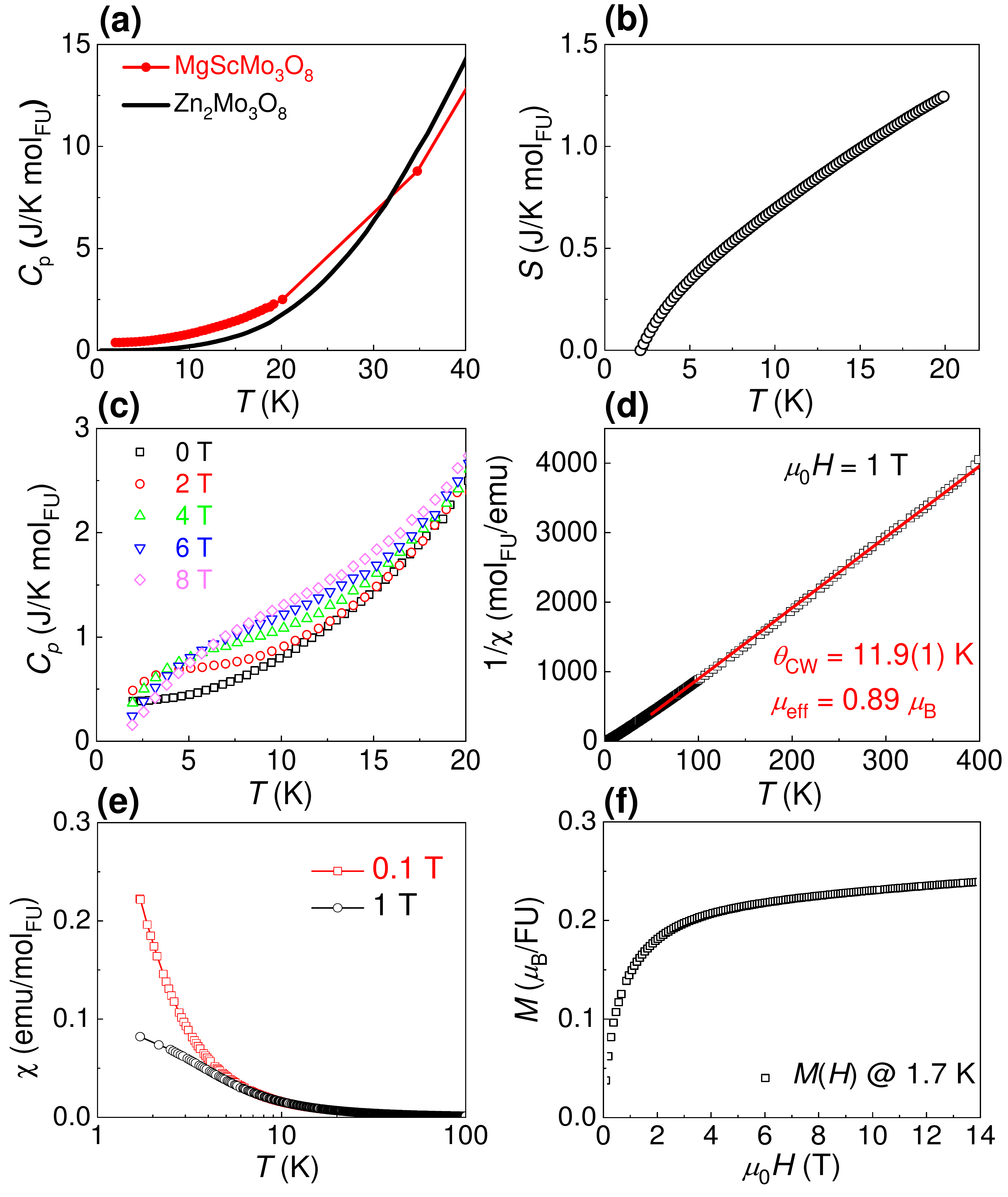}
\end{center}
\par
\caption{(color online) For MgScMo$_3$O$_8$, 
(a) the $T$-dependence of the zero-field specific heat, along with data for the non-magnetic analog Zn$_2$Mo$_3$O$_8$;
(b) the magnetic entropy released when warming to 20 K;
(c) the $T$-dependence of the specific heat for selected magnetic fields;
(d) the $T$-dependence of the inverse dc susceptibility and the associated Curie-Weiss fit (solid line);
(e) the $T$-dependence of the dc susceptibility for selected fields over the low-$T$ region only;
(f) the magnetization $M$ vs applied field $H$ at $T =$~1.7~K. }
\label{Fig7}
\end{figure}

We measured the $T$-dependence of the specific heat for MgScMo$_3$O$_8$, with the raw zero-field data shown in Fig.~\ref{Fig7}(a). Notably, the data levels off at the lowest temperatures and does not seem to extrapolate to zero as $T \rightarrow$ 0. This behavior may be indicative of a magnetic transition just below the 1.8~K base temperature of our measurement. Although the Zn$_2$Mo$_3$O$_8$ data doesn't provide a perfect lattice match, MgScMo$_3$O$_8$ appears to be a low-dimensional and/or frustrated magnet as the magnetic contribution in the heat capacity seems to persist up to temperatures well-above a possible magnetic transition. We isolated the magnetic signal using the same procedure described for ZnScMo$_3$O$_8$ above, with the magnetic entropy recovered upon warming to 20~K presented in Fig.~\ref{Fig7}(b). Again, there is a significant reduction in the recovered entropy expected for a spin-1/2 magnetic phase transition. Although this reduction may arise from a combination of magnetic frustration and low-dimensionality, as discussed above, we expect some missing entropy simply due to not cooling the sample well below the magnetic transition temperature. Our specific heat data in selected applied fields, illustrated in Fig.~\ref{Fig7}(c), provides further evidence for the zero-field ordered ground state. A broad peak, similar to the one observed for ZnScMo$_3$O$_8$, is identified here and again it shifts to higher temperatures with increasing field. These results suggest that MgScMo$_3$O$_8$ hosts a zero-field ordered ground state with a net moment and a transition temperature below 1.8~K.

The dc susceptibility ($\chi = M/H$) for MgScMo$_3$O$_8$, plotted in Fig.~\ref{Fig7}(d) and (e) as 1/$\chi$ and $\chi$ respectively, is also indicative of an ordered state with a net moment. A Curie-Weiss fit of the high-temperature regime (between 50 and 400 K) for 1/$\chi$ yields a Curie-Weiss temperature $\theta_{\text{CW}}$ = 11.9(1)~K and an effective moment $\mu_{\text{eff}}$ = 0.89(3)~$\mu_{\text{B}}$/trimer. The positive Curie-Weiss temperature again signifies dominant ferromagnetic interactions and the lower magnitude compared to the value for ZnScMo$_3$O$_8$ is consistent with a suppressed magnetic transition temperature. The reduced effective moment, relative to the Zn analog, may be at least partially explained by the significant non-magnetic impurity fraction in this sample, although it is still consistent with each [Mo$_3$]$^{11+}$ trimer unit hosting a $S =$~1/2 degree of freedom. The low-field (i.e. 0.1~T) $\chi$ data obeys the Curie-Weiss law down to the base temperature of 1.8~K, which was expected based on the 0~T heat capacity data. On the other hand, the 1~T susceptibility data shows deviations from the Curie-Weiss law at low temperatures with the concave down behavior that is expected for a transition to an ordered state with a net moment, with a $T_c =$~3~K identified using the minimum in $d\chi/dT$. This value agrees well with the field-evolution of the magnetic transition temperature observed in the specific heat measurements. The magnetization vs field at 1.7~K, as shown in Fig.~\ref{Fig7}(f), reveals a net moment that saturates at 0.24~$\mu_B$/trimer. Again, the significantly reduced value as compared to expectations for a spin-1/2 ferromagnet may arise from the close proximity of the measurement to the magnetic transition temperature, the importance of unquenched orbital angular momentum to the ground state, and/or the realization of a canted antiferromagnetic ground state. 

\begin{figure*}
\centering
\includegraphics[width=5.4in]{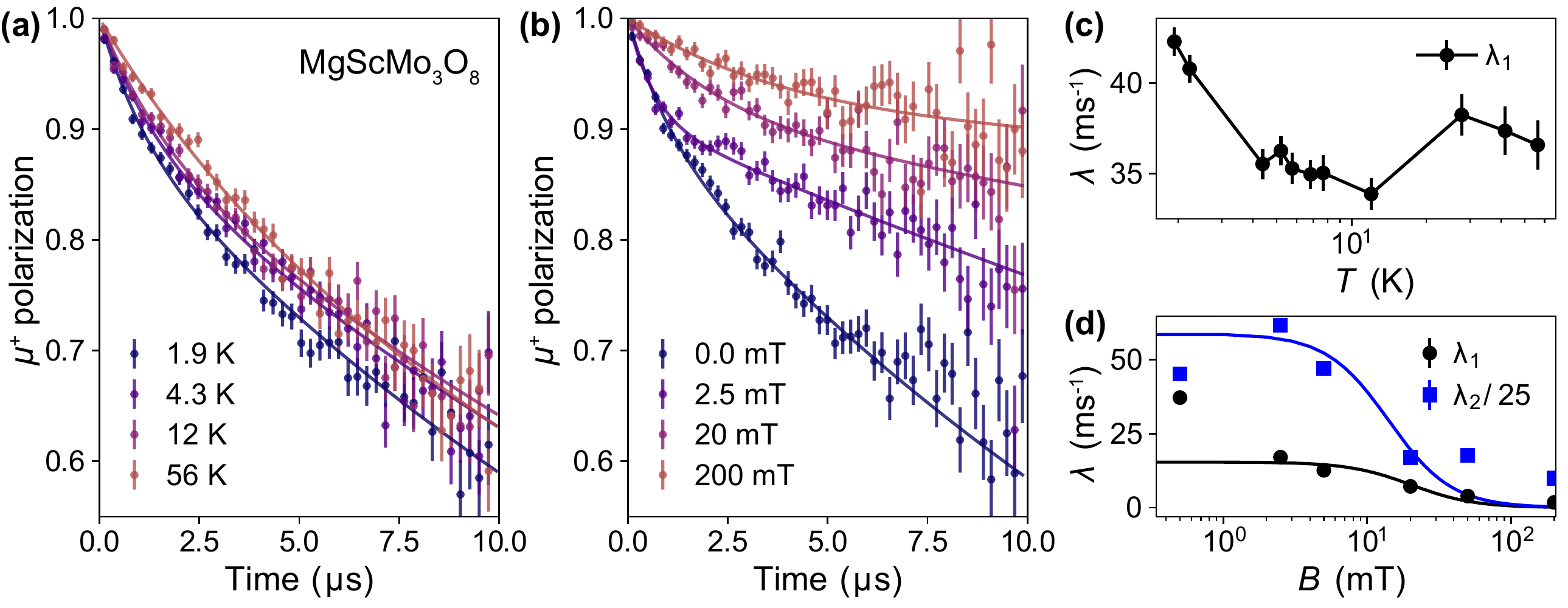}
\caption{(color online) For MgScMo$_3$O$_8$, (a) muon spin polarization vs time for MgScMo$_3$O$_8$ in zero field at selected temperatures. (b) The longitudinal field dependence (decoupling) of the muon spin polarization in MgScMo$_3$O$_8$. The solid lines are again double-exponential fits. (c) Temperature dependence of the dominant (slow-relaxing) signal. (d) Field dependence of the relaxation rates of both components fit with Redfield theory (Eq.~\ref{Redfield}) for fields 5 mT and higher.
\label{Fig8}}
\end{figure*}

As with the other samples, we performed zero- and longitudinal-field $\mu$SR measurements to assess the magnetic ground state of MgScMo$_3$O$_8$. ZF-$\mu$SR spectra at selected temperatures are shown in Fig.~\ref{Fig8}(a). In sharp contrast to the Zn analog, this sample shows very weak relaxation at all temperatures studied with the zero-field muon spin relaxation changing very little as a function of temperature. Again, two relaxation components appear to be necessary to cleanly fit the data, especially in a longitudinal magnetic field, as shown in Fig.~\ref{Fig8}(b). Once again, we have fit the data using a double exponential function (Eq.~\ref{DoubleExponential}) with a constant amplitude ratio. In this material, the faster-relaxing component only accounts for about 9\% of the signal. Some of the relaxation at low magnetic fields is undoubtedly a result of weak nuclear magnetism which should result in a Gaussian Kubo-Toyabe, however with such weak relaxation, we found it impossible to separate the nuclear and electronic contributions. In Fig.~\ref{Fig8}(c) we show the temperature dependence of the relaxation rate for the dominant ($\sim 91$\%) component, that is $\lambda_1(T)$. There is very little variation in this value, indicating a lack of magnetic ordering in the temperature range studied. We suspect that the given statistical error bars underestimate the uncertainty on $\lambda_1$ and that there may be other systematic sources of error. 

In Fig.~\ref{Fig8}(d) we show the field dependence of the relaxation of both components: $\lambda_1(B)$ for the majority fraction and $\lambda_2(B)$ for the minority fraction. For longitudinal fields below 5~mT we expect that much of the relaxation originates from weak nuclear fields, hence we have excluded these data points from the fitting with Redfield theory (Eq.~\ref{Redfield}) which is shown with the solid curves in Fig.~\ref{Fig8}(d). For the majority component $\lambda_1$, we obtain a field of $\Delta_{d1} = 0.43(0.05)$~mT that is fluctuating relatively slowly, with $\omega_{c1} = 17(5)$~MHz. For the minority fraction $\lambda_2$, the fitting parameters are $\Delta_{d2} = 3.5(0.6)$ mT and $\omega_{c2} = 12(5)$~MHz. Given the small amplitude of the signal and the poor quality of the fit, it is difficult to draw any firm conclusions regarding this component, though it at least seems likely that it is dynamic in origin. If it were static, it would represent a very small internal field of about 1.0 mT. However, the field scan in Fig.~\ref{Fig8}(b) shows that it is not entirely decoupled at 20 mT, nor have we been able to find any evidence in x-ray diffraction for a $\sim 9$\% impurity phase that might exhibit static magnetism. This component could possibly arise from defects or vacancies within the sample that lead to Mo sites with a larger magnetic moment than the bulk of the sample. One likely scenario is that some Mg has been introduced onto the Mo sites due to their similar ionic radii and the large amount of excess MgO that was required in the synthesis process. 

Regardless of what gives rise to the minority fraction, our zero-field $\mu$SR results show that MgScMo$_3$O$_8$ remains in a paramagnetic state down to temperatures as low as 1.93 K, despite the much larger Weiss constant of $\theta_\mathrm{CW} = 11.9$ K, indicative of some source of frustration in the system and/or a highly two-dimensional character of the interactions. 

\subsection{Phase Diagram}

\begin{figure}[tbp]
\linespread{1}
\par
\begin{center}
\includegraphics[width=\columnwidth]{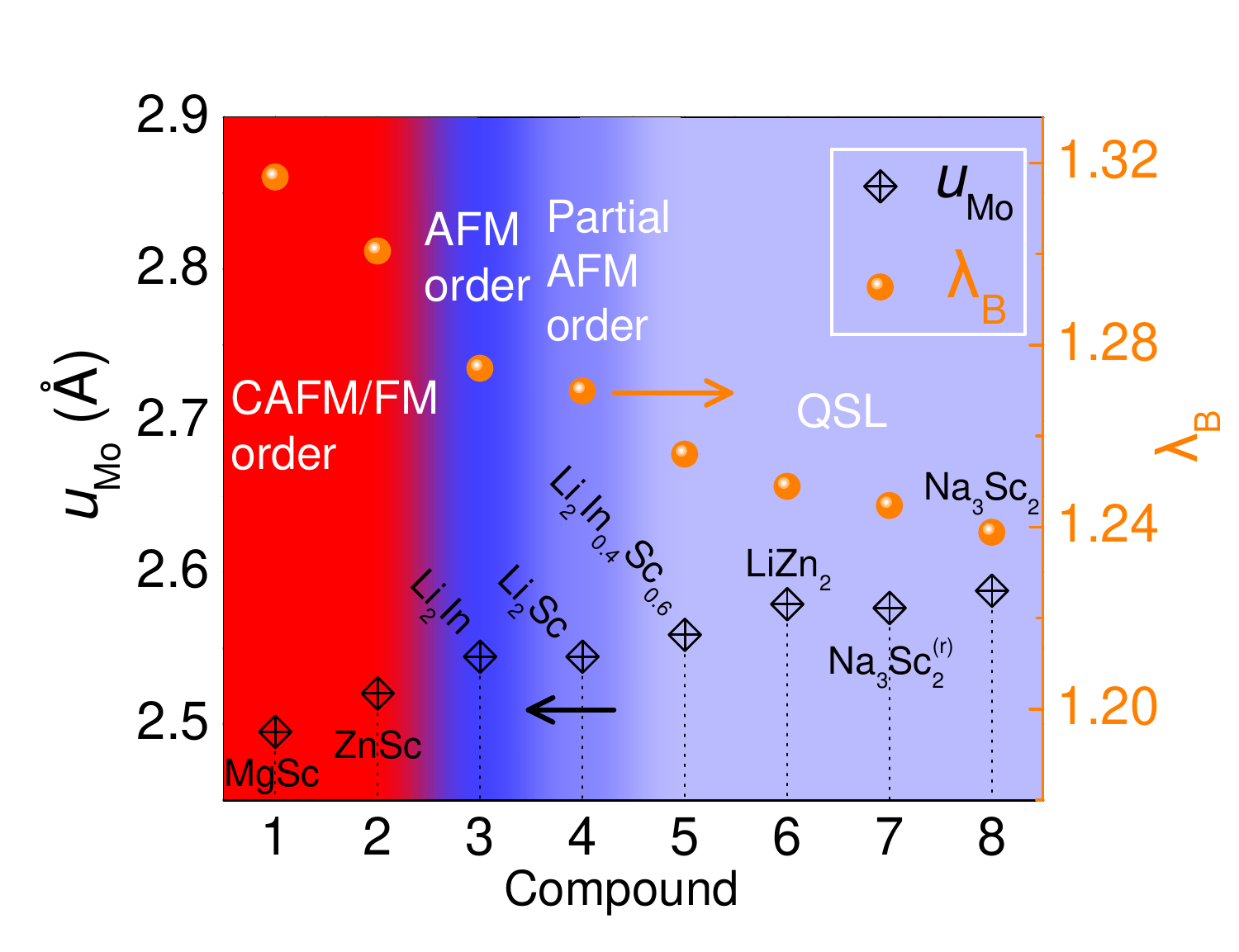}
\end{center}
\par
\caption{(color online) Phase diagram showing the evolution of the magnetic ground states for the Mo trimer molecular magnets as a function of the nearest neighbor Mo-Mo distance $u_{Mo}$ (left axis) and the breathing parameter $\lambda_B$ (right axis). The materials included are MgScMo$_3$O$_8$ (MgSc, current work), ZnScMo$_3$O$_8$ (ZnSc, current work), Li$_2$InMo$_3$O$_8$ (Li$_2$In, \cite{15_haraguchi}), Li$_2$ScMo$_3$O$_8$ (Li2Sc, \cite{15_haraguchi}), Li$_2$In$_{0.4}$Sc$_{0.6}$Mo$_3$O$_8$ (Li$_2$In$_{0.4}$Sc$_{0.6}$, \cite{18_akbari}), LiZn$_2$Mo$_3$O$_8$ (LiZn$_2$, \cite{12_sheckelton}), the previously reported Na$_3$Sc$_2$Mo$_5$O$_{16}$ (Na$_3$Sc$_2^{(r)}$, \cite{17_haraguchi_2}), and our Na$_3$Sc$_2$Mo$_5$O$_{16}$ sample (Na$_3$Sc$_2$, current work). The acronyms have the following meanings: FM (ferromagnetism), CAFM (canted antiferromagnetism), AFM (antiferromagnetism), and QSL (quantum spin liquid). 
}
\label{Fig9}
\end{figure}

With the magnetic ground state of Na$_3$Sc$_2$Mo$_5$O$_{16}$ confirmed and the magnetic ground states of ZnScMo$_3$O$_8$ and MgScMo$_3$O$_8$ now established, we present a comprehensive phase diagram for the Mo trimer family in Fig.~\ref{Fig9}. More specifically, we show how their magnetic ground states evolve as a function of both the nearest neighbor Mo-Mo distance $u_{Mo}$ and the breathing parameter $\lambda_B$. For simplicity, the chemical compositions in the phase diagram are abbreviated to include the first two elements only (e.g. MgScMo$_3$O$_8$ is written as MgSc). Our main finding is that magnetically-ordered ground states are realized when $u_{Mo}$ is small and $\lambda_B$ is large, while quantum spin liquid candidates are generated in the opposite limit. Li$_2$ScMo$_3$O$_8$ \cite{18_akbari, 15_haraguchi, 19_iida}, with intermediate values for both parameters, hosts short-range order and therefore represents an interesting crossover case that warrants further investigation.   

The results presented in our phase diagram are well-explained by previous theoretical work \cite{16_chen, 21_nikolaev} based on an extended Hubbard model with hopping parameters $t_u$, $t_d$ for the up and down triangles on the breathing kagome lattice, intersite Coulomb interactions $V_u$, $V_d$, and an on-site Coulomb interaction $U$. In the strong interaction limit where $|t_u| < V_u$ and $|t_d| < V_d$ that is expected for large $u_{Mo}$ and small $\lambda_B$, plaquette charge order has been identified. Each hexagon of the breathing kagome lattice is decorated by three unpaired electrons that occupy nonadjacent sites, with next-nearest neighbor hexagons forming resonant states consisting of one valence bond and one unpaired spin. This unique charge order can be stabilized by nearest-neighbor antiferromagnetic exchange interactions \cite{16_chen} or hopping parameters with opposite signs \cite{21_nikolaev}. Since the plaquette charge order represents a specific type of resonating valence bond state, Mo trimer systems in the strong interaction limit are quantum spin liquid candidates. 

When $t_u$ becomes comparable to $V_d$, as expected for smaller $u_{Mo}$ and larger $\lambda_B$ values, the unpaired electrons minimize their energy by forming molecular bound states leading to cluster Mott insulating behavior. When both $t_u$ and $t_d$ are large, collective ground states with long-range magnetic order can arise on the $S =$~1/2 triangular lattice formed by the Mo trimers. In fact, the 120$^\circ$ long-range antiferromagnetic order for Li$_2$InMo$_3$O$_8$ \cite{19_iida} is expected for the NN Heisenberg model with this lattice geometry. The ordered ground states that we identify here for ZnScMo$_3$O$_8$ and MgScMo$_3$O$_8$ are likely due to NN exchange interactions that are ferromagnetic in these cases instead. We also speculate that the lower magnetic transition temperature for MgScMo$_3$O$_8$, as compared to its Zn analog, can be attributed to weaker NN exchange interactions arising from the larger $d_{Mo}$ and $\lambda_B$ values. 

\section{IV. CONCLUSIONS}
In this work, we determined the crystal structures of the three [Mo$_3$]$^{11+}$-based molecular magnets \ch{Na3Sc2Mo5O16}, \ch{ZnScMo3O8}, and \ch{MgScMo3O8} using x-ray or neutron powder diffraction. We also investigated the magnetic properties of these materials with ac \& dc susceptibility, specific heat, neutron powder diffraction, and $\mu$SR measurements. We find that \ch{ZnScMo3O8} and \ch{MgScMo3O8} host ordered ground states with net moments (low-moment ferromagnetism or canted antiferromagnetism) and \ch{Na3Sc2Mo5O16} is a quantum spin liquid candidate. By drawing on previous work, we also constructed a phase diagram for the [Mo$_3$]$^{11+}$-based molecular magnets that illustrates the high tunability of their magnetic ground states with only small changes in the nearest neighbor Mo-Mo distance and the breathing parameter. Our results are well-explained in the context of an extended Hubbard model, and they show that molecular magnets based on 4d-transition metals are an excellent playground for identifying and characterizing exotic states of matter and novel magnetic phenomena arising from competing energy scales \cite{16_streltsov, 20_chen, 21_khomskii}. 

\begin{acknowledgments}
We acknowledge valuable conversations with S. Streltsov and Y. B. Kim as well as technical support of the CMMS team at TRIUMF, in particular B. Hitti, G. Morris, D. Arseneau and D. Vyas. Q.C., R.S., and H.D.Z. thank the support from NSF-DMR-2003117. The work of Z.D. and M.M. at Georgia Tech was supported by NSF-DMR-1750186. J.Q. and A.A.S. acknowledge funding from the Natural Sciences and Engineering Research Council and the Canada First Excellence Research Fund. A portion of this research used resources at the High Flux Isotope Reactor, which is a DOE Office of Science User Facility operated by Oak Ridge National Laboratory. A portion of this work was performed at the NHMFL, which is supported by National Science Foundation Cooperative Agreement No. DMR-1157490 and the State of Florida. 
\end{acknowledgments}

\end{document}